\documentclass[a4paper,twocolumn,english,8pt,amssymb,nofootinbib,superscriptaddress]{revtex4}
\usepackage{lmodern}
\usepackage{lmodern}

\usepackage[T1]{fontenc}
\usepackage[latin9]{inputenc}
\usepackage{babel}
\usepackage{mathrsfs}
\usepackage{amsmath}
\usepackage{amssymb}
\usepackage{esint}
\usepackage[unicode=true,pdfusetitle,
 bookmarks=true,bookmarksnumbered=false,bookmarksopen=false,
 breaklinks=false,pdfborder={0 0 1},backref=false,colorlinks=false]
 {hyperref}

\makeatletter

\@ifundefined{textcolor}{}
{%
 \definecolor{BLACK}{gray}{0}
 \definecolor{WHITE}{gray}{1}
 \definecolor{RED}{rgb}{1,0,0}
 \definecolor{GREEN}{rgb}{0,1,0}
 \definecolor{BLUE}{rgb}{0,0,1}
 \definecolor{CYAN}{cmyk}{1,0,0,0}
 \definecolor{MAGENTA}{cmyk}{0,1,0,0}
 \definecolor{YELLOW}{cmyk}{0,0,1,0}
}

\usepackage{babel}
\usepackage{babel}
\usepackage{babel}
\usepackage{babel}
\usepackage{babel}
\usepackage{babel}
\usepackage{babel}
\usepackage{babel}
\usepackage{babel}
\usepackage{babel}
\usepackage{babel}
\usepackage{babel}
\usepackage{babel}
\usepackage{babel}
\usepackage{babel}
\usepackage{babel}

\usepackage{babel}
\usepackage{graphicx}
\def\b{\begin{equation}}
\def\e{\end{equation}}

\@ifundefined{textcolor}{}{%
 \definecolor{BLACK}{gray}{0}
 \definecolor{WHITE}{gray}{1}
 \definecolor{RED}{rgb}{1,0,0}
 \definecolor{GREEN}{rgb}{0,1,0}
 \definecolor{BLUE}{rgb}{0,0,1}
 \definecolor{CYAN}{cmyk}{1,0,0,0}
 \definecolor{MAGENTA}{cmyk}{0,1,0,0}
 \definecolor{YELLOW}{cmyk}{0,0,1,0}
 }

\usepackage{latexsym}\usepackage{bm}

\makeatother

\begin{document}
\title{ Exact formulas for spherical photon orbits around Kerr black holes }

\author{Aydin Tavlayan}

\email{aydint@metu.edu.tr}

\selectlanguage{english}%

\affiliation{Department of Physics,\\
 Middle East Technical University, 06800 Ankara, Turkey}
\author{Bayram Tekin}

\email{btekin@metu.edu.tr}
\affiliation{Department of Physics,\\
 Middle East Technical University, 06800 Ankara, Turkey}

\selectlanguage{english}%

\begin{abstract}
\noindent Exact formulas relating the radii of the spherical photon orbits to the
black hole's rotation parameter and the effective inclination angle of the orbit have been known only for equatorial and polar orbits up to now. Here we provide exact formulas for nonequatorial orbits that lie between these extreme limits. For a given rotation parameter of the black hole, there is a critical inclination angle below which there are four null photon orbits two of which are in the exterior region. At the critical angle, the radii of these orbits are explicitly found. We also provide approximate but very accurate  formulas for any orbit. To arrive at these analytical solutions, we carefully study the sextic polynomial in the radius that arises for null spherical geodesics. 
\end{abstract}
\maketitle

\section{{\normalsize{}{}{}{}{}{}{}{}{}{}{}{}INTRODUCTION}} Observational black hole physics, using gravitational waves or radio waves, is currently thrilling: since the first black hole merger observation via gravitational waves \cite{merger}, there has been more than a dozen observations, including the most recent groundbreaking detection of a binary black hole system with a total mass of $150 ~ M_{\odot}$ \cite{intermediate}. With radio waves, the Event Horizon Telescope provided us with a stunning image of the environment and the shadow of a supermassive black hole \cite{ET}.  All these black holes are described by the  Kerr solution of the vacuum Einstein equation with mass $m$ and angular momentum $J$ \cite{Kerr} and no other parameters. Hence, studying physics, especially the motion of ultrarelativistic particles and light around the Kerr black hole metric, is extremely relevant to the current observations, especially in the context of imaging the black hole and the ring-down phase of a merger. 

Among the light orbits around a rotating black hole, there is a particular class with constant radii that are astrophysically relevant  \cite{Cunha}. For this class of light orbits, there are exact expressions of their radii in terms of the black hole rotation or spin parameter and the effective inclination angle of the orbit for only two special cases. These are the equatorial orbits (called the light rings) and the polar orbit, but in between these two extremes, called the spherical photon orbits \cite{Teo1}, there are no known exact solutions in the literature.  Here we provide exact expressions for an important class of such orbits that are not restricted to the equatorial or polar planes. For these orbits, there is a relation between the inclination angle and the black hole rotation parameter. Moreover, this angle turns out to be critical in the sense that there are four null geodesics below it, albeit two of these are inside the event horizon for nonextremal black holes; one of the external orbits has a nonmonotonic dependence on the rotation parameter of the black hole in contrast to the monotonic dependence of the polar and equatorial orbits. On the other hand, above the critical inclination angle, there are only two spherical null geodesics. We shall also give approximate analytical expressions not only for the critical case but also for the case of generic  inclination angle. 

\section{{\normalsize{}{}{}{}{}{}{}{}{}{}{}{}NULL GEODESICS}} The metric of the Kerr black hole with mass $m$ and angular momentum $J = m a$ in the Boyer-Lindquist coordinates (in the $G= c=1$ units) is 
\begin{eqnarray}\label{BL}
ds^2=-\Big(1-{{2 m r}\over{\Sigma}}\Big)dt^2-{{4m ar\sin^2\theta}\over{\Sigma}}dt
d\phi+{{\Sigma}\over{\Delta}}dr^2 \nonumber \\
+\Sigma\, d\theta^2+\Big(r^2+a^2+{{2m a^2r\sin^2\theta}\over{\Sigma}}\Big)\sin^2\theta
d\varphi^2,
\end{eqnarray}
where $a$ is called the rotation parameter which can be taken to be positive without  loss of generality, and 
\begin{equation}
 \Delta\equiv r^2-2mr+a^2,\hskip 1 cm 
\Sigma\equiv r^2+a^2\cos^2\theta.
\end{equation}
The larger root of $\Delta =0$  is the event horizon located at
\begin{equation}
r_{\text{H}}=m+(m^2-a^2)^{1/2}.
\end{equation}
For the spherical photon orbits, the relevant part of the geodesic equations that we shall be interested in is the radial one \cite{Carter, Bardeen,Wilkins,Teo1},
\begin{equation}
\Sigma \,{{dr}\over{d\lambda}}=\pm\sqrt{R(r)},
\end{equation}
where $\lambda$ is an affine parameter along the null geodesics and the radial function is given as
\begin{eqnarray}
R(r)&\equiv&E^2 r^4 + (a^2 E^2 - L_z^2 - {\cal Q})r^2 \nonumber \\
&&+ 2 m \Big ( ( a E - L_z)^2 +{\cal Q} \Big)r- a^2 {\cal Q} . \label{Rfunction}
\end{eqnarray}
Here $E$ is the conserved energy of the photons (related with the $\xi_{(t)} = \frac{\partial}{\partial t}$ Killing vector of the metric), $L_z$ is the conserved $z$-component of the angular momentum of the photons (related  to the $\xi_{(\varphi)} = \frac{\partial}{\partial \varphi}$ Killing vector), and  ${\cal Q}$ is the Carter's constant related with a symmetric rank two Killing tensor whose explicit form is not needed here; it suffices to know that  it is connected with the motion perpendicular to the equatorial plane and  ${\cal Q} \ge 0 $ for constant radii orbits, satisfying the bound for equatorial orbits.

Two equations must be satisfied, $R(r)=0$ and $\frac{ d R(r)}{dr}=0$, 
for constant radius $r$ null geodesics. These yield the following (physically viable) equations \cite{Teo1}:
\begin{eqnarray}
&&\frac{L_z}{E} = - \frac{ r^3-3 m r^2+a^2 r+a^2 m}{a (r-m)}, \label{ang}\\ 
&&\frac{{\cal Q}}{E^2} = -\frac{r^3 \left(r^3 -6 m r^2 +9 m^2 r -4 a^2 m \right)}{a^2 (r-m)^2}.\label{Carter}
\end{eqnarray}
Defining the {\it effective inclination angle} as \cite{Ryan}
\begin{equation}
\cos i \equiv \frac{L_z}{\sqrt{ L_z^2 + {\cal Q}}},
\end{equation}
which is clearly a conserved quantity, and noting the fact that photons with different energies can orbit at the same radius as dictated by the equivalence principle,
one can eliminate the energy of the photon in (\ref{ang}, \ref{Carter}) and end up with the following sextic polynomial equation:
\begin{eqnarray}
p(x)&\equiv &x^6-6 x^5+(9+2 \nu u) x^4-4 u x^3-\nu  u (6 -u) x^2 \nonumber \\
&&+2 \nu  u^2 x +\nu  u^2=0,
\label{sextic}
\end{eqnarray}
where we defined the dimensionless variables
\begin{equation}
 x \equiv \frac{r}{m}, \hskip 1 cm u \equiv \frac{a^2}{m^2}, \hskip 1 cm \nu \equiv  \sin^2 i.\hskip 1 cm  
\end{equation}
From now on, we will call $u$ to be the rotation parameter.
So, one must solve this polynomial equation as $x = x(u,\nu)$  for the following intervals:
\begin{equation}
 0< x ,\hskip 1 cm   0 \le \nu  \le 1,\hskip 1 cm 0 \le u\le 1. \label{interval}
\end{equation}
Let us first briefly discuss the known solutions in the particular cases before we present novel solutions.

\section{{\normalsize{}{}{}{}{}{}{}{}{}{}{}{}EQUATORIAL AND POLAR ORBITS}}
To study the equatorial orbits (${\cal Q} \equiv 0$) , let $ i = 0$ or $\pi$. Then $\nu =0$ and the nontrivial part of the sextic (\ref{sextic}) reduces to the cubic
\begin{equation}
x^3-6 x^2+9 x-4 u =0,
\end{equation}
of which the discriminant is ${\cal D} = 4 u (-1 + u)$. For $u=1$, which is the extremal black hole case, there is a double root at $x=1$ and another root at $x=4$. The double root has the same $r$-coordinate  as the event horizon, but they are not located at the same point as the horizon due to the extended geometry of the extremal black hole's throat. This tricky case was addressed in \cite{Bardeen, Jacobson}. For the subextremal case $u <1$ and so ${\cal D} <0$: there are three distinct real roots,
\begin{eqnarray}
&& x_\pm = 2 +2\cos \left(\frac{2}{3} \cos ^{-1}(\pm \sqrt{u}  )\right), \label{retro-pro}  \\
&& x_{\text{in}}= 4 \sin ^2\left(\frac{1}{3} \sin ^{-1}\sqrt{u}\right),
\label{retro-pro}
\end{eqnarray}
where $x_+$ is the retrograde orbit satisfying $  3 \le x_+  \le 4$, while $x_{-}$ is the prograde orbit satisfying $  1 \le x_- \le 3$. The third orbit is inside the event horizon: 
$ 0 < x_{\text{in}} \le 1$.

For the polar orbit ($L_z= 0$), and so $ i =\pm \frac{\pi}{2}$, then $\nu =1$ and (\ref{sextic})  reduces to
\begin{equation}
x^3-3 x^2+u x +u=0.
\end{equation}
There are three real solutions, but one of the solutions has $x <0$ and hence it is unphysical.
The external solution is
\begin{equation}
x_P= 1+2 \sqrt{1-\frac{u}{3}} \cos \left \{\frac{1}{3} \cos ^{-1}\left(\frac{1-u}{\left(1-\frac{u}{3}\right)^{3/2}}\right)\right \},
\end{equation}
which lies  in the region $ 1 \le x_P \le 3$. The interior solution is
\begin{equation}
x_{\text{in}}=1-2 \sqrt{1-\frac{u}{3}} \sin \left \{\frac{1}{3} \sin ^{-1}\left(\frac{1-u}{\left(1-\frac{u}{3}\right)^{3/2}}\right)\right \},
\end{equation}
which lies  in the region $ 0 <x_{\text{in}} \le 1$. The photons in these polar orbits have zero angular momentum, but they do rotate with the black hole as a result of the frame-dragging (or the Lense-Thirring) effect.

\section{{\normalsize{}{}{}{}{}{}{}{}{}{}{}{}SCHWARZSCHILD BLACK HOLE}}
 For $u=0$, the case for which the Kerr black hole has no rotation and reduces to the Schwarzschild black hole,  the nontrivial part of  (\ref{sextic}) for generic $\nu$ simplifies to  
\begin{equation}
x-3= 0,
\end{equation}
which is the radius of the circular photon ring (or the light ring) independent of the inclination angle due to the spherical symmetry.

\section{{\normalsize{}{}{}{}{}{}{}{}{}{}{}{}SPHERICAL PHOTON ORBITS}} 
For generic $\nu$ and $u$, there are four sign variations in the polynomial (\ref{sextic}), and hence Descartes' rule of sign change says that there can be 4, 2, or 0 positive roots.
 Using a  Tschirnhaus transformation of the form $y=\alpha x^4+\beta x^3+\gamma x^2+\delta x+\epsilon$, we can bring the sextic to the reduced form $y^6+c_1 y^2+c_2 y+c_3=0$, albeit with very complicated coefficients $c_i = c_i (u,\nu)$. The exact expressions of the roots in the generic case of any polynomial beyond the quartic, in terms of a finite number of radicals, are not possible as dictated by the celebrated Abel-Ruffini impossibility theorem. This theorem applies  to our sextic in the original or reduced form. However, one can certainly find exact formulas for the roots in terms of special functions such as the Mellin formula \cite{Mellin}, theta functions \cite{theta}, or the nested radicals. Because of the complexity of exact roots in the most general form, let us concentrate on particular values of $\nu$ and $u$ of the sextic for which we can find exact solutions in radicals. Remarkably, this will turn out to be possible. 

As we are looking for the real roots of the sextic, we might search for the point in the ($u,\nu$) parameter space where four real and two complex roots arise as allowed by the Descartes' rule of sign.  It turns out for a given rotation parameter $u$, there is a {\it critical inclination angle} below which there are four real roots and above which there are two real roots. For this critical value,  $\nu_{\text{cr}}$= $\nu_{\text{cr}}(u)$, the discriminant of the sextic vanishes, and one can find all the roots of the sextic in an exact form. 

To proceed further, it pays to define the following variables which will simplify the final expressions:
\begin{equation}
\nu \equiv \frac{\xi}{u}, \hskip 1 cm u \equiv 1- w^3,
\end{equation}
with $ 0 \le \xi \le 1$ and $0 \le w \le 1$. Then it can be shown that the sextic (\ref{sextic}) reduces to a quadratic times a quartic at the following critical point:
\begin{equation}
\xi_{\text{cr}}=\frac{3 (1-w)^3}{ 7+(w-5) w}.
\label{critical}
\end{equation}
The domain of $\xi_{\text{cr}}$ is $[0,\frac{3}{7}]$. Note that previously in \cite{Hod}, one particular point in this critical domain was discovered: namely, it was shown that the critical angle   corresponding to the extremal black hole ($u=1$) is  $\cos i= \sqrt{4/7}$ or $\nu = \frac{3}{7}$.  Here with the formula (\ref{critical}) the critical angle for {\it any} rotating parameter is given. So, at this critical angle, the sextic factors as 
\begin{equation}
p(x)= (w+x-1)^2 q(x),
\end{equation}
with the quartic part given as 
\begin{eqnarray}
q(x)=x^4 +\alpha x^3 + \beta x^2 + \gamma x+ \sigma,
\end{eqnarray}
where the coefficients are 
\begin{eqnarray}
&&\alpha=-2 (w+2) \nonumber \\
&&\beta=\frac{3w \Big (w^2(w-5)+3 w+8\Big)+6 }{(w-5) w+7} \nonumber \\
&&\gamma=\frac{6 (w-2) (w-1) \left(w^2+w+1\right) }{(w-5) w+7}\nonumber \\
&&\sigma=\frac{3 (w-1)^2 \left(w^2+w+1\right)}{(w-5) w+7}.
\end{eqnarray}

The quadratic part has a double root at
\begin{equation}
x = 1- ( 1- u)^{\frac{1}{3}}
\end{equation}
inside the event horizon for finite $u$ and approaches the event horizon as $u \rightarrow 1$. The quartic part has generically two real and two complex roots, which can be written in explicit form, but the expressions are rather cumbersome, and so we do not depict them here. These explicit formulas can be used, with the help of a computer, to understand the properties of the orbits exactly. Here we give several such computations.,

In Fig. 1, we plot the real roots of the sextic as a function of the rotation parameter at the critical inclination angle. One can see how the roots are  connected with each other and observe the nonmonotonic behavior of the retrograde orbit. Note that, as opposed to the critical retrograde orbit, the equatorial retrograde orbit is monotonic in the rotation parameter; all the other roots are also monotonic. In Fig. 2, we plot the dimensionless angular momentum per energy, $\frac{L_z}{ m E}$, as given in (\ref{ang}) as a function of $u$.  In Fig. 3, we plot $ \frac{d^2 R}{ d x^2}$ as a function of $u$ and show that the critical null geodesics are unstable in the radial direction as expected: all spherical photon orbits are unstable in the radial direction; this fact makes them very interesting as they can escape to infinity due to a small perturbation, and this is part of the light detected by the Event Horizon Telescope. In Fig. 4, we plot the dimensionless impact parameter for the retrograde orbits at the critical inclination angle and at the equator. Among all the orbits, the retrograde orbit has the highest impact parameter. Figure 5 shows the analogous plot for the prograde orbits.

\begin{figure}[ht!]
	\centering
	\includegraphics[width=1\linewidth]{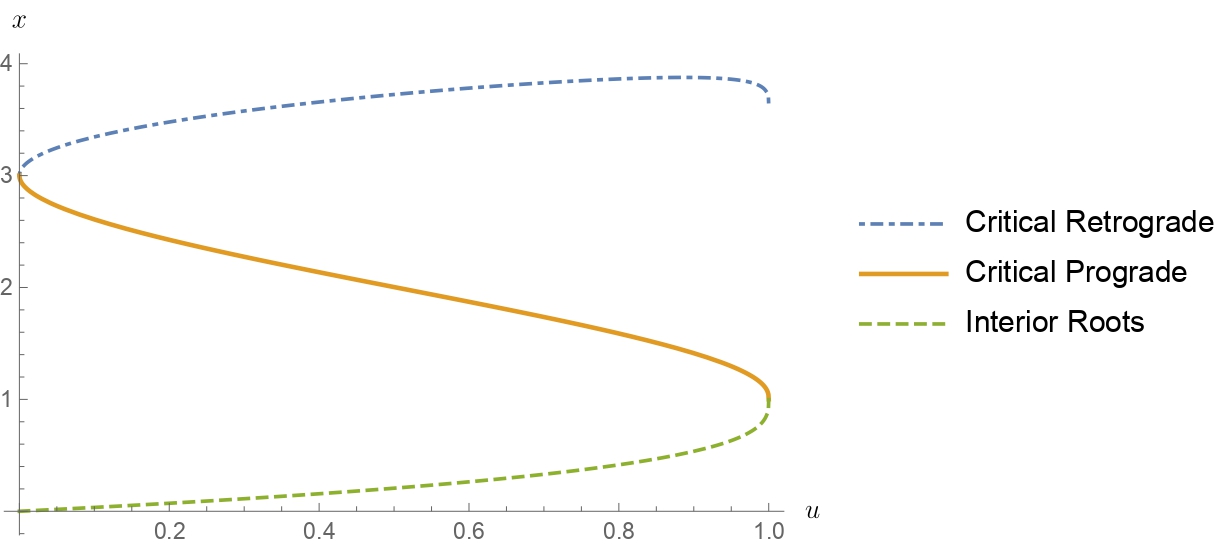}
	\caption{Real roots of the sextic are plotted as a function of the rotation parameter at the critical inclination angle. The root  corresponding to the retrograde orbit  is nonmonotonic in $u$ while the other roots are monotonic. Note that the third and the fourth roots are double roots and lie inside the event horizon for the nonextremal  black hole.
 }
\end{figure}

\begin{figure}[ht!]
	\centering
	\includegraphics[width=1\linewidth]{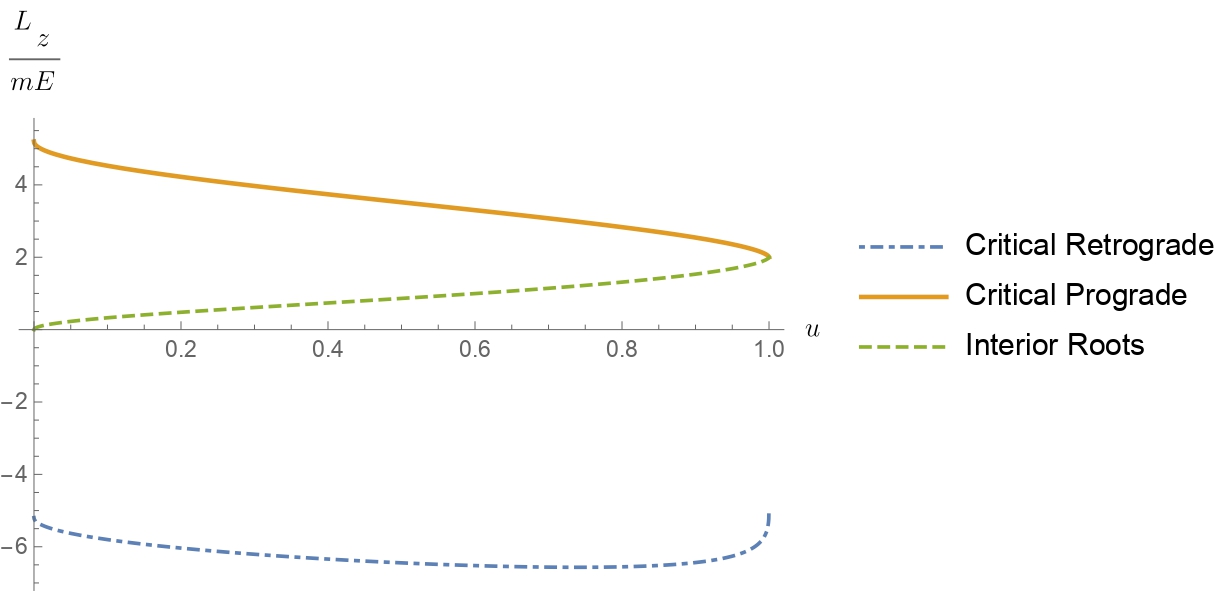}
	\caption{The $z$-component of the angular momentum per unit energy and black hole mass is plotted for the real roots at the critical point as a function of the rotation parameter.  The root  corresponding to the retrograde orbit with a negative angular momentum component has a nonmonotonic dependence on $u$. 
 }
\end{figure}

\begin{figure}[h!]
	\centering
	\includegraphics[width=1\linewidth]{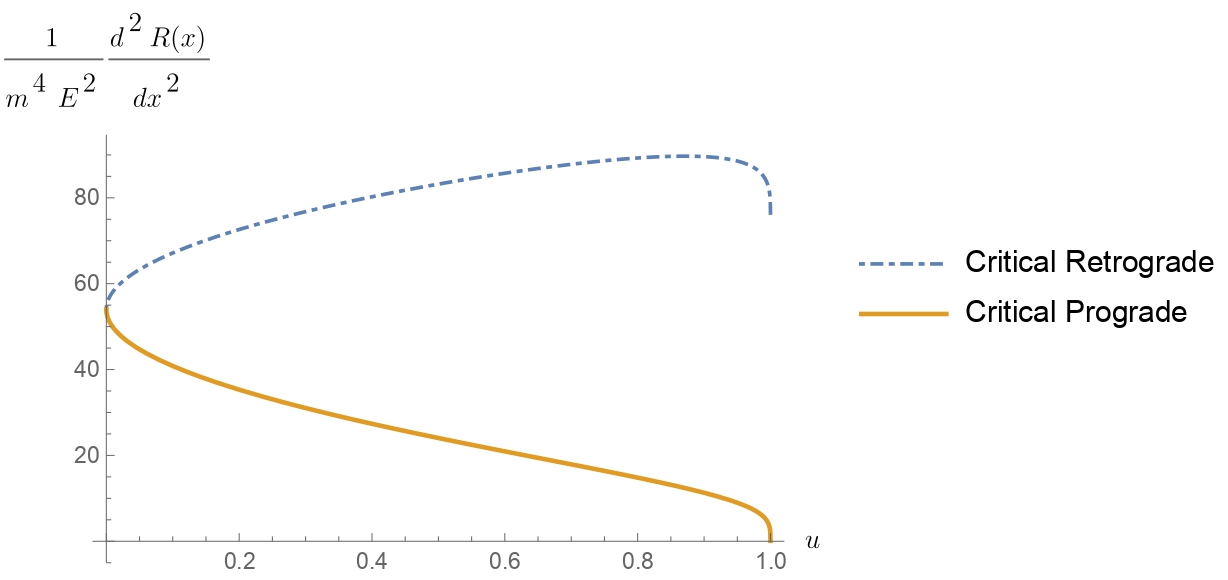}
	\caption{Second derivative of  $R$ from (\ref{Rfunction}) is given for the critical photon orbits: for the exterior roots, the second derivative is positive and hence they are unstable orbits just like the prograde, retrograde, and the polar orbits. For the interior double root, the second derivative vanishes, so that root is a saddle point.
 }
\end{figure}
\begin{figure}[h!]
	\centering
	\includegraphics[width=1\linewidth]{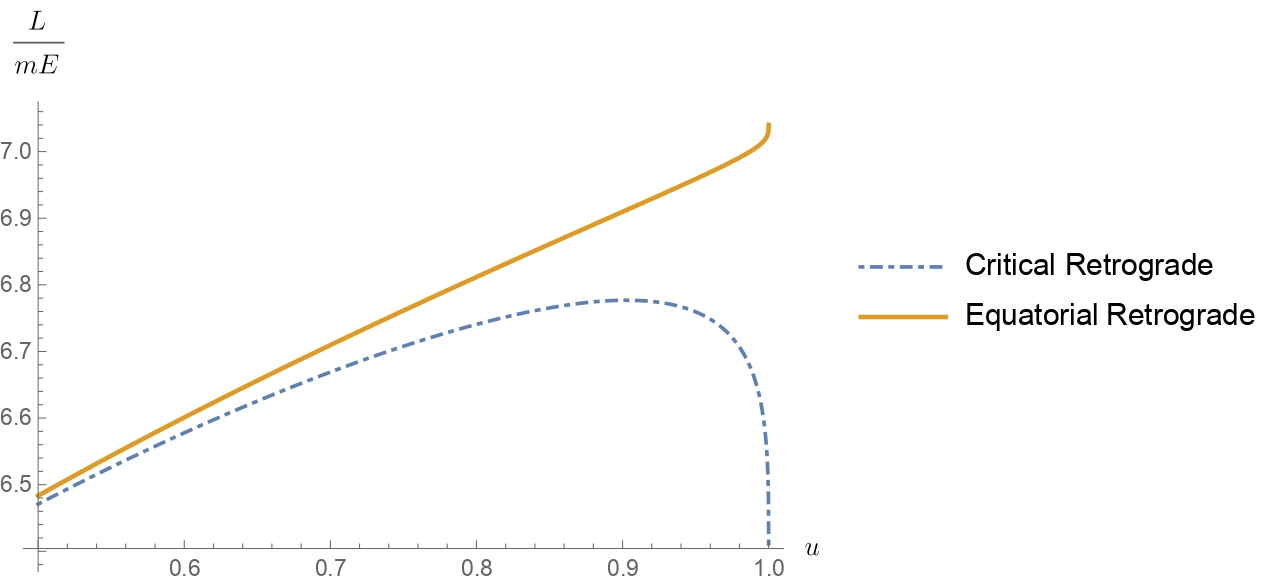}
	\caption{ The dimensionless ''impact parameter'' $\frac{L}{ m E}$ with $L = \sqrt{L_z^2 + {\mathcal{Q}}}$ for the retrograde orbits at the critical inclination angle and at the equatorial plane ,s plotted. 
The equatorial retrograde orbit  has the highest impact parameter and is monotonic in the rotation parameter, while the impact parameter of the critical retrograde  orbit is nonmonotonic. 
 }
\end{figure}

\begin{figure}[h!]
	\centering
	\includegraphics[width=1\linewidth]{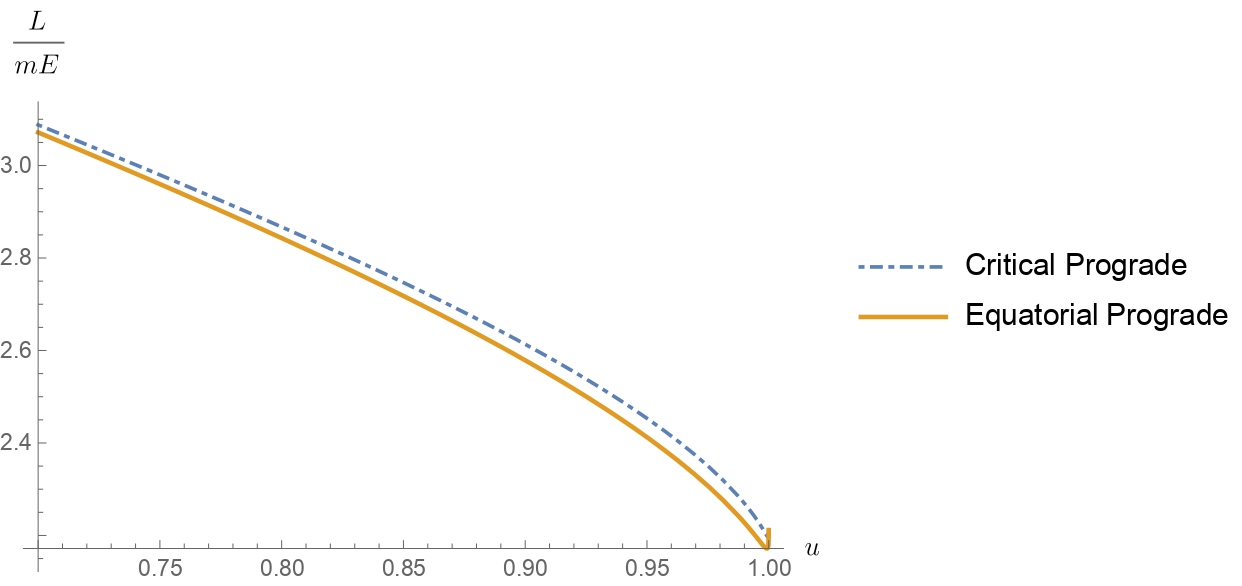}
	\caption{ The dimensionless impact parameter for the critical and equatorial prograde orbits are nonmonotonic in the impact parameter and they are both smaller than those of retrograde orbits.
 }
\end{figure}

\section{{\normalsize{}{}{}{}{}{}{}{}{}{}{}{}APPROXIMATE SOLUTIONS}} Here we present a method to develop approximate solutions around any of the known solutions presented above. The method is based on the  Lagrange-B\"{u}rmann  inversion theorem \cite{Henrici}. Let us assume $f$ to be a function of the form 
\begin{equation}
f(x)=\nu, \hskip 1 cm \frac{df(x)}{dx}\Bigr|_{\substack{x=x_0}} \neq 0.
\end{equation}
Then the Lagrange-B\"{u}rmann inversion theorem states that one can write $x$ as a power series  in $\nu$ of the form
\begin{equation}
x=x_0+\sum_{n>0} \frac{g_n}{n!}\Big(\nu-f(x_0)\Big)^n,
\end{equation}
where 
\begin{equation}
g_n=\lim_{x \to x_0} \frac{d^{n-1}}{dx^{n-1}}\left(\frac{x-x_0}{f(x)-f(x_0)}\right)^n, \;\;\; n=1,2,...\;. \label{coeff}
\end{equation}

To apply this method to the sextic, for a given $u$, one can solve for $\nu$ in (\ref{sextic}) to get the inclination angle as a function of the radius as 
\begin{equation}
\nu(x)=-\frac{x^3 \Big((x-3)^2 x-4 u\Big)}{u \Big(u (x+1)^2+2 \left(x^2-3\right) x^2\Big)}.
\end{equation}
For $x_0$, as stated above, one can choose any of the known solutions. Here, for the sake of simplicity, let us choose the prograde and the retrograde orbits  (\ref{retro-pro}) as examples.  Then $x_0= x_\pm$ and  $\nu(x_0)=0$; one can easily compute the approximate solution up to the desired order. Here we shall keep up to ${\mathcal {O}}(\nu)$ as it already yields an accurate approximation. Then, from (\ref{coeff}), one finds
\begin{eqnarray}
g_1=\lim_{x \to x_0} \left(\frac{x-x_0}{\nu}\right), \nonumber 
\end{eqnarray}
which yields
\begin{eqnarray}
&&g_1=\frac{A(x_0)}{B(x_0)}\\
&&A(x_0)=u \Big (u (x_0+1)^2+2x_0^4-6 x_0^2\Big)^2\nonumber\\
&&B(x_0)=4 x_0^2 (x_0+3) \Big(u (x_0+1)+x_0^3-3 x_0^2\Big)\nonumber\\
&&\hskip 1.3 cm \times \Big(u-x_0^3 +3 x_0^2-3 x_0)\Big).\nonumber
\end{eqnarray}
Then, the resulting approximate solution is
\begin{equation}
x=x_0+g_1 \nu+{\mathcal {O}}(\nu^2),
\end{equation}
which is accurate for small $\nu \approx 0.1$ by two parts in $10^4$ for slowly rotating black holes, $u \approx 0.1 $, and by two parts in $10^3$ for fast rotating ones, $u \approx 0.9$,  around the prograde equatorial orbits. The approximation works better for retrograde orbits. 
\begin{table}[h!]
	\centering
	\caption{Comparison of approximate and exact numerical results around the prograde orbit at the equator, $x_0=x_+$. Approximation is done with the  Lagrange-B\"{u}rmann expansion up to order ${\mathcal {O}}(\nu^5)$.}
	\begin{tabular}{|c|c|c|c|c|} 
		\hline
		$\hskip 0.3 cm  \nu \hskip 0.3 cm $ & $ \hskip 0.3 cm u \hskip 0.3 cm$ & $ \hskip 0.3 cm  x_{\text{approximate}} \hskip 0.3 cm$   & $ \hskip 0.3 cm x_{\text{exact}}$         & \hskip 0.3 cm \text{Error} (\%)  \hskip 0.3 cm       \\ 
		\hline
		0.1 & 0.1 & 2.6247277 & 2.6247278 & $4.3\times10^{-6}$  \\ 
		\hline
		0.1 & 0.5 & 2.027428 & 2.027429 & $1.4 \times10^{-5}$    \\ 
		\hline
		0.1 & 0.9 & 1.411893   & 1.411895 & $7.5\times10^{-5}$     \\ 
		\hline
		0.5 & 0.1 & 2.7033  & 2.7038 &        $1.9\times10^{-2}$  \\ 
		\hline
		0.5 & 0.5 & 2.174 & 2.175 &  $6 \times 10^{-2}$       \\ 
		\hline
		0.5 & 0.9 & 1.562 & 1.565 &   $2.4\times10^{-1}$      \\
		\hline
	\end{tabular}
\end{table}
\begin{table}[h!]
	\centering
	\caption{Comparison of approximate and exact numerical results around the retrograde orbit at the equator, $x_0=x_-$. Approximation is done with the  Lagrange-B\"{u}rmann expansion up to order ${\mathcal {O}}(\nu^5)$.}
	\begin{tabular}{|c|c|c|c|c|} 
		\hline
		$\hskip 0.3 cm  \nu \hskip 0.3 cm $ & $ \hskip 0.3 cm u \hskip 0.3 cm$ & $ \hskip 0.3 cm  x_{\text{approximate}} \hskip 0.3 cm$   & $ \hskip 0.3 cm x_{\text{exact}}$         & \hskip 0.3 cm \text{Error} (\%)  \hskip 0.3 cm       \\
		\hline
		0.1 & 0.1 & 3.3249272  & 3.3249271 & $3.3\times10^{-6}$  \\ 
		\hline
		0.1 & 0.5 & 3.6801618 & 3.6801615 & $7.6\times10^{-6}$      \\ 
		\hline
		0.1 & 0.9 & 3.8803947  & 3.8803943 & $1.1\times10^{-5}$     \\ 
		\hline
		0.5 & 0.1 & 3.228  & 3.227 & $1.6\times10^{-2}$         \\ 
		\hline
		0.5 & 0.5 & 3.438 & 3.437 & $3.8\times10^{-2}$          \\ 
		\hline
		0.5 & 0.9 & 3.533 & 3.531 & $5.6\times10^{-2}$        \\
		\hline
	\end{tabular}
\end{table}

For better accuracy, one needs to compute higher-order terms. We show some examples up to and including  ${\mathcal{O}}(\nu^4) $ in the tables.  As the errors show, the approximation is exceptionally accurate. In Table I and II, the exact numerical values are compared with the approximate analytical ones obtained via the Lagrange-B\"{u}rmann expansion, respectively,  around the prograde and retrograde orbits at the equatorial plane.  Near $\nu =0$, around which the series was expanded, the error is less than one part in $10^{4}$ for all of the rotation parameter values . The error becomes larger for larger values of $\nu$: for example, for $\nu =0.5$, the error is $0.24 \%$ for highly spinning black holes.  Hence, one can use these approximate analytical formulas to explore the behavior of these orbits.

\section{{\normalsize{}{}{}{}{}{}{}{}{}{}{}{}CONCLUSIONS}}  The motion of light in the vicinity of a Kerr black hole is important for astrophysics. Up to now, the radii of the polar and equatorial plane light orbits have been known explicitly in terms of the rotation parameter of the black hole. Here we have extended the known analytical solutions by showing that there are  two exterior nonequatorial light orbits whose radii can be found analytically for a given black hole with a rotation parameter $a$. One of these orbits is a prograde orbit, which monotonically decreases in the rotation parameter and the other one is a retrograde orbit, which is nonmonotonic, in contrast to the equatorial retrograde orbit that is monotonic. This feature was observed in \cite{Hod} for the extremal black hole case.  The effective inclination angle of these two orbits is critical in the sense that exactly at and below that angle, there are always four spherical null orbits.
Two of these are inside the event horizon, and at the critical angle, these two orbits merge into a single orbit  with a radius, which can be analytically expressed in terms of the rotation parameter. 
These two interior exact solutions, which are presumably not significant for astrophysics, may  nevertheless  be important to better explore the geometry of the black hole region. The novel solutions presented here are all unstable in the radial direction, as all the spherical geodesics are unstable; this fact is closely related to the stability and the shadow of the Kerr black hole \cite{Cunha2}. An extended discussion of the material here and the discussion for the timelike geodesics along the similar lines will be given elsewhere \cite{AT}.

\section*{ACKNOWLEDGMENTS} We would like to thank Ferit \"{O}ktem, Atalay Karasu, \c{C}etin \"{U}rti\c{s}, and \"{O}zg\"{u}r Ki\c{s}isel for useful discussions.

\end{document}